\documentclass[twocolumn,showpacs,superscriptaddress,amsmath,amssymb,aps,pra]{revtex4-1}
\usepackage{bm}
\usepackage{graphicx}
\usepackage{dcolumn}
\usepackage{amsmath,txfonts}
\usepackage{epstopdf}
\usepackage[mathlines]{lineno}
\usepackage{color}
\usepackage[colorlinks=true,linkcolor=blue,citecolor=blue]{hyperref}
\usepackage[normalem]{ulem}
\usepackage{makecell}
\usepackage{tabularx}

\begin{document}

\title{Nonreciprocal superradiant quantum phase transition induced by the magnon Kerr effect}

\author{Guo-Qiang Zhang}
\email{zhangguoqiang@hznu.edu.cn}
\affiliation{School of Physics, Hangzhou Normal University, Hangzhou, Zhejiang 311121, China}

\author{Si-Yan Lin}
\affiliation{School of Physics, Hangzhou Normal University, Hangzhou, Zhejiang 311121, China}

\author{Wei Feng}
\affiliation{School of Physics, Hangzhou Normal University, Hangzhou, Zhejiang 311121, China}

\author{Lijiong Shen}
\affiliation{School of Physics, Hangzhou Normal University, Hangzhou, Zhejiang 311121, China}

\author{Yi-Hao Kang}
\affiliation{School of Physics, Hangzhou Normal University, Hangzhou, Zhejiang 311121, China}

\author{Wei Xiong}
\email{xiongweiphys@wzu.edu.cn}
\affiliation{Department of Physics, Wenzhou University, Zhejiang 325035, China}
\affiliation{International Quantum Academy, Shenzhen, 518048, China}

\begin{abstract}
Recently, proposals for realizing a nonreciprocal superradiant quantum phase transition (SQPT) have been put forward, based on either nonreciprocal interactions between two spin ensembles or the Sagnac-Fizeau shift in a spinning cavity. However, experimental implementation of such a nonreciprocal SQPT remains challenging. This motivates the search for new mechanisms capable of producing a nonreciprocal SQPT. Here, we propose an alternative approach to realize a nonreciprocal SQPT, induced by the magnon Kerr effect (MKE), in a cavity magnonic system, where magnons in a yttrium iron garnet (YIG) sphere are coupled to cavity photons. The MKE coefficient is positive ($K>0$) when the bias magnetic field is aligned along the crystallographic axis [100], but negative ($K<0$) when aligned along the axis [110]. We show that the steady-state phase diagram for $K > 0$ differs markedly from that for $K < 0$. This contrast is the origin of the nonreciprocal SQPT. By further studying the steady-state magnon occupation and its fluctuations versus the parametric drive strength, we demonstrate that the SQPT becomes nonreciprocal, characterized by distinct critical thresholds for $K > 0$ and $K < 0$. Moreover, we introduce a bidirectional contrast ratio to quantify this nonreciprocal behavior. Our work provides a new mechanism for realizing the nonreciprocal SQPT, with potential applications in designing nonreciprocal quantum devices.
\end{abstract}

\date{\today}

\maketitle

\section{Introduction}

Superradiant quantum phase transition (SQPT) in the Dicke model was first predicted by Hepp {\it et al.} in 1973~\cite{Hepp73,Wang73}. The Dicke model describes the collective coupling of $N$ two-level atoms to a quantized light field~\cite{Dicke54}. In the thermodynamic limit with the number of atoms $N \rightarrow +\infty$, the system undergoes a SQPT from the normal phase to the superradiant phase at zero temperature as the coupling strength exceeds a critical value~\cite{Emary03PRL,Emary03PRE}. Below the critical coupling, the system is in the normal phase, where the atoms remain in their ground state and the cavity field is in the vacuum state. Above the critical coupling, the system enters the superradiant phase, characterized by collective excitations of both the atomic ensemble and the cavity field. With advances in experimental capabilities, the SQPT in the Dicke model and its variants (e.g., Tavis-Cummings model~\cite{Castanos09,Zou13}, Rabi model~\cite{Ashhab13,Hwang15}, Jaynes-Cummings model~\cite{Hwang16,Liu24}, and coupled multi-oscillator model~\cite{Zhang21-Chen,Qin22,Wang24Nori}) has been observed in a variety of quantum platforms. These include Bose-Einstein condensates within optical cavities~\cite{Baumann10,Baumann11,Brennecke13,Baden14}, degenerate Fermi gases inside an optical cavity~\cite{Zhang21Chen,Wu23}, circuit QED systems~\cite{Feng15,Zheng23,Ning24}, trapped ions~\cite{Cai21,Zhao25}, nuclear-magnetic-resonance systems~\cite{Chen21,Wu24}, and so on.

Recently, Fruchart {\it et al.} established a general theory of nonreciprocal phase transitions, showing that time-dependent phases arise from asymmetric interactions among multiple species or fields~\cite{Fruchart21}. The nonreciprocal phenomenon refers to that the system responds differently when an external field is applied in opposite directions~\cite{Barzanjeh25}. Due to potential applications in quantum technologies, various nonreciprocal quantum effects have been studied, such as thermal noise routers~\cite{Barzanjeh18}, nonreciprocal magnonic diodes~\cite{Zou24,Nakata24,Yuan23}, one-way quantum amplifiers~\cite{Metelmann15,Malz18,Shen18}, and nonreciprocal photon (or magnon) blockade~\cite{Huang18,Wang22}. Following the framework of Ref.~\cite{Fruchart21}, the nonreciprocal SQPT was proposed in a nonreciprocal Dicke model, with quantum-light-mediated nonreciprocal interactions between two spin ensembles~\cite{Chiacchio23}. In a parallel approach, two groups have independently proposed realizing a nonreciprocal SQPT by exploiting the Sagnac-Fizeau shift in a spinning cavity~\cite{Zhu24,Xu24}. However, the key elements of these proposals---specifically, the required nonreciprocal interactions between two spin ensembles~\cite{Chiacchio23} and the Sagnac-Fizeau shift in a spinning cavity~\cite{Zhu24,Xu24}---present major challenges for existing experimental quantum platforms that have demonstrated the SQPT~\cite{Baumann10,Baumann11,Brennecke13,Baden14,Zhang21Chen,Wu23,Feng15,Zheng23,Ning24,Cai21,Zhao25,Chen21,Wu24}. It is therefore imperative to explore new theoretical mechanisms for realizing a nonreciprocal SQPT.

Owing to their excellent tunability and design flexibility, cavity magnonic systems have emerged as well-suited platforms for exploring novel physical phenomena~\cite{Soykal10,Tabuchi14,Zhang14,Goryachev14,Zhang15,Zhang15Zou,Bai15,Lachance-Quirion19,Rameshti22,Yuan22}. In Ref.~\cite{Wang16}, frequency shifts of both the magnon and cavity modes induced by the magnon Kerr effect (MKE) were experimentally observed by directly driving a yttrium iron garnet (YIG) sphere within a microwave cavity. The MKE, originating from magnetocrystalline anisotropy in YIG, describes magnon-magnon interactions. It can lead to multistability of cavity magnon polaritons~\cite{Wang18,Zhang19,Nair20,Bi21,Shen21,Shen22}, magnon-photon entanglement~\cite{Zhang19Scully,Yang21Jin}, long-distance spin-spin coupling~\cite{Xiong22Tian}, and unidirectional microwave transmission~\cite{Kong19,Ullah24,Miao24}, among other effects. A notable feature of the MKE is that its coefficient depends on the direction of the bias magnetic field applied to the YIG sphere: the MKE coefficient is positive when the field is aligned with the crystallographic axis [100], but negative when aligned with the axis [110]~\cite{Wang18,Zhang19}. It is this directional dependence that makes cavity magnonic systems a promising platform for studying nonreciprocal quantum effects, such as nonreciprocal quantum entanglement~\cite{Chen23,Chen24,Ahmed24,Kong24,Liu25,Kong25}, nonreciprocal magnon (or photon) blockade~\cite{Zhang24,Fan24}, and nonreciprocal quantum synchronization~\cite{Lai25}.

In this work, we propose a scheme to realize a nonreciprocal SQPT induced by MKE in a cavity magnonic system. The system consists of a magnon mode in a YIG sphere coupled to a cavity mode via the magnetic-dipole interaction, where the cavity is subject to a parametric drive, and the magnon mode exhibits the MKE. By analyzing the stability of the magnon occupation solutions, we obtain the steady-state phase diagram, which comprises three distinct phases: normal phase, superradiant phase and bistable phase. The phase diagram differs significantly when the bias magnetic field is aligned along the [100] and [110] crystallographic axes of the YIG sphere, corresponding to positive ($K>0$) and negative ($K<0$) MKE coefficients, respectively~\cite{Wang18,Zhang19}. We further examine the steady-state magnon number and its fluctuations as functions of the parametric drive strength, revealing that the SQPT becomes nonreciprocal, with different critical thresholds for $K>0$ and $K<0$. To quantify this behavior, we introduce a bidirectional contrast ratio that characterizes the degree of nonreciprocity in the SQPT. Our work provides an alternative approach for realizing a nonreciprocal SQPT~\cite{Fruchart21,Chiacchio23,Zhu24,Xu24}.

The remainder of this paper is structured as follows. In Sec.~\ref{model}, we introduce the proposed cavity magnonic system and derive the steady-state solutions for the magnon occupation. Section~\ref{NSQPT} presents the steady-state phase diagram based on a stability analysis of these solutions, followed by a detailed investigation of the MKE-induced nonreciprocal SQPT. The feasibility of experimentally implementing the proposed scheme is discussed in Sec.~\ref{experimental}. Finally, discussions and conclusions are given in Sec.~\ref{conclusions}.

\section{Model}\label{model}

\begin{figure}
\includegraphics[width=0.45\textwidth]{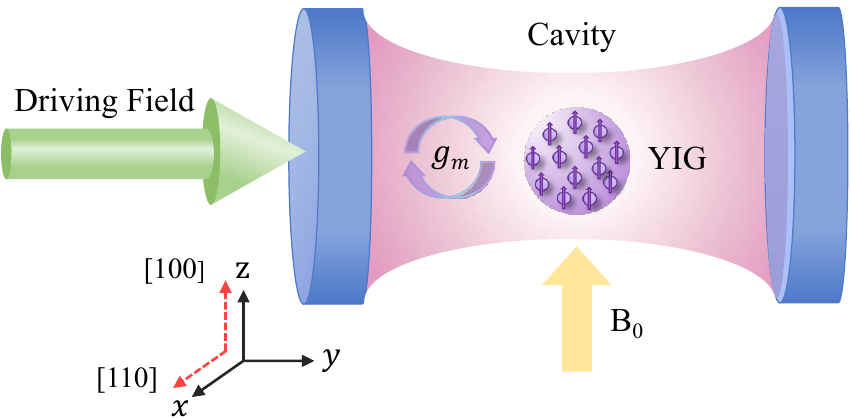}
\caption{Schematic diagram of the proposed cavity magnonic system. The system comprises a parametrically driven microwave cavity and a YIG sphere magnetized to saturation by a bias magnetic field $B_0$. The magnetic field is aligned along either the crystallographic axis [100] or [110] (see the red-arrowed lines) of the YIG sphere.}
\label{fig1}
\end{figure}

As depicted in Fig.~\ref{fig1}, the considered cavity magnonic system consists of a microwave cavity and a millimeter-sized YIG sphere, where the cavity is pumped by a parametric drive, and the YIG sphere is magnetized to saturation by a bias static magnetic field. The YIG sphere supports a Kittel mode (with frequency $\omega_m$), which is coupled to the cavity mode (with frequency $\omega_a$) via the collective magnetic-dipole interaction. Taking into account both the system's dissipation and the interaction among magnons (i.e., MKE), the parametrically driven cavity magnonic system can be described by the following effective non-Hermitian Hamiltonian (see Appendix \ref{Appendix-A}; hereafter setting $\hbar=1$):
\begin{eqnarray}\label{NHH}
H_{\rm eff}&=&(\Delta_{a}-i\kappa_a)a^{\dag}a+(\Delta_{m}-i\gamma_m)m^{\dag}m+\frac{K}{2}m^{\dag}m^{\dag}mm\nonumber\\
           & & +g_m(a^{\dag}m+am^{\dag})+\frac{\Omega}{2}(a^{\dag}a^{\dag}+aa),
\end{eqnarray}
where $a^{\dag}$ and $a$ ( $m^{\dag}$ and $m$) are the creation and annihilation operators of the cavity mode (magnon mode), $\Delta_{a(m)}=\omega_{a(m)}-\omega_{d}/2$ $(>0)$ is the frequency detuning of the cavity mode (magnon mode) from the drive field,
$\kappa_a$ ($\gamma_m$) is the decay rate of the cavity mode (magnon mode), $K$ is the MKE coefficient, $g_m$ is the coupling strength between cavity mode and magnon mode, and $\omega_{d}$ ($\Omega$) is the frequency (strength) of parametric drive. Here the MKE results from the magnetocrystalline anisotropy of the YIG sphere, and the sign of MKE coefficient $K$ can be controlled by adjusting the direction of the bias magnetic field~\cite{Wang18,Zhang19}. When the bias magnetic field is aligned along the crystallographic axis [100], the MKE coefficient $K$ is positive ($K>0$). In contrast, $K$ becomes negative ($K<0$) for aligning the bias magnetic field along the crystallographic axis [110]. The dependence of $K$ on the direction of the bias magnetic field is the origin of the nonreciprocal SQPT (cf.~Sec.~\ref{NSQPT}). The effective Hamiltonian $H_{\rm eff}$ maintains invariance under the transformation $a \rightarrow -a$ and $m \rightarrow -m$, demonstrating parity symmetry~\cite{Emary03PRL,Emary03PRE}. Defining the parity operator $\Pi=\exp[i\pi(a^\dag a+m^\dag m)]$, this symmetry can be expressed as the commutation relation $[H_{\rm eff},\Pi]=0$. In our system, the spontaneous breaking of parity symmetry occurs when the cavity magnonic system undergoes a phase transition from the normal phase to superradiant phase.

\renewcommand\tabcolsep{4.4pt}
\begin{table}
	\centering
	\scriptsize
	\caption{The parameter conditions (PCs) for $|M|^2_{\pm}>0$ in both cases of $K>0$ and $K<0$, where the solution $|M|^2=|M|^2_-$ ($|M|^2=|M|^2_+$) is unstable for all parameter values in the case of $K>0$ ($K<0$) and not considered here.}
	\label{tab:notations}
    \small
	\begin{tabular}{ccc}
        \hline
		 \makecell[cc]{nonnegativity\\ constraint}
                                    & \makecell[cc]{PCs for $K>0$} & \makecell[cc]{PCs for $K<0$}\\
		\hline
		\makecell[cc]{$|M|^2_+>0$}  &\makecell[cc]{$\Omega>\Omega_1$ for $\Delta_{m}/\Delta_{a}<\xi$;\\
                                    $\Omega>\Omega_2$ for $\Delta_{m}/\Delta_{a}>\xi$}     & \makecell[cc]{trivial}\\
        \hline
		\makecell[cc]{$|M|^2_->0$}  & \makecell[cc]{trivial} &
                                 \makecell[cc]{$\Omega>\Omega_2$ for $\Delta_{m}/\Delta_{a}<\xi$;\\
                                   $\Omega>\Omega_1$ for $\Delta_{m}/\Delta_{a}>\xi$} \\
		\hline
	\end{tabular}
\label{Table1}
\end{table}

The dynamics of the cavity magnonic system can be described through the following quantum Langevin equations~\cite{Walls94}:
\begin{eqnarray}\label{Langevin}
\dot{a}&=&-i(\Delta_{a}-i\kappa_a)a-ig_m m-i\Omega a^{\dag}+\sqrt{2\kappa_a}\,a_{\rm in},\nonumber\\
\dot{m}&=&-i(\Delta_{m}-i\gamma_m)m-iKm^{\dag}mm-ig_m a+\sqrt{2\gamma_m}\,m_{\rm in},
\end{eqnarray}
where $o_{\rm in}$ is the input noise operator for the mode $o$ ($o=a,\, m$), satisfying $\langle o_{\rm in} \rangle=0$.
Under the Markovian approximation, the noise operator $o_{\rm in}$ is characterized by the following two-time correlation
functions: $\langle o_{\rm in}(t) o_{\rm in}^\dag (t') \rangle=[\overline{n}_o(\omega_o) +1]\delta(t-t')$ and $\langle o_{\rm in}^\dag(t) o_{\rm in} (t') \rangle=\overline{n}_o(\omega_o) \delta(t-t')$ with $\overline{n}_o(\omega_o)=[\exp(\hbar \omega_o/{k_B T})-1]^{-1}$ denoting the equilibrium mean thermal excitation number of the mode $o$, where $k_B$ is the Boltzmann constant and $T$ is the bath temperature. To analyze the steady-state behavior of the cavity magnonic system, we perform a linearization of Eq.~(\ref{Langevin}) by decomposing the operators as $a = A + \delta a$ and $m = M + \delta m$, where $A$ ($M$) represents the expectation value of operator $a$ ($m$), and $\delta a$ ($\delta m$) denotes the quantum fluctuation with $\langle \delta a\rangle = \langle \delta m\rangle = 0$. Following from Eq.~(\ref{Langevin}), the expectation values $A$ and $M$ satisfy
\begin{eqnarray}\label{expected-value}
\dot{A}&=&-i(\Delta_{a}-i\kappa_a)A -ig_mM-i\Omega A^{*},\nonumber\\
\dot{M}&=&-i(\Delta_{m}+K|M|^2-i\gamma_m)M-ig_m A,
\end{eqnarray}
and the fluctuation operators $\delta a$ and $\delta m$ obey
\begin{eqnarray}\label{fluctuations}
\delta\dot{a}&=&-i(\Delta_{a}-i\kappa_a)\delta a-ig_m\delta m-i\Omega\delta a^{\dag}+\sqrt{2\kappa_a}a_{\rm{in}},\nonumber\\
\delta\dot{m}&=&-i(\widetilde{\Delta}_{m}-i\gamma_m)\delta m-ig_m\delta a-iF\delta m^{\dag}+\sqrt{2\gamma_m}m_{\rm{in}},
\end{eqnarray}
where $\widetilde{\Delta}_{m}=\Delta_{m}+2K|M|^2$ and $F=KM^{2}$. For obtaining the linear fluctuation equation (\ref{fluctuations}), all small high-order fluctuation terms have been neglected. The steady-state solutions for $|M|^2$ and $|A|^2$, obtained from Eq.~(\ref{expected-value}) with $\dot{A}=\dot{M}=0$, exhibit three branches due to the MKE:
\begin{eqnarray}\label{solutions}
|M|^{2}_0=0,~~~
|M|^{2}_{\pm}=\Big(-\Delta'_{m} \pm \sqrt{\eta^{2}\Omega^{2}-{\gamma'_{m}}^{2}}\Big)/K,
\end{eqnarray}
and
\begin{eqnarray}\label{}
|A|^{2}_0=0,~~~
|A|^{2}_{\pm}=\left[\big(\Delta_m+K|M|_{\pm}^2\big)^2+\gamma_m^2\right]|M|_{\pm}^2/g_m^2,
\end{eqnarray}
where $\Delta'_{m}=\Delta_{m}-\eta\Delta_{a}$, $\gamma'_{m}= \gamma_{m}+\eta\kappa_{a}$ and $\eta=g_{m}^{2}/(\Delta_{a}^{2}+\kappa_{a}^{2}-\Omega^{2})$. Obviously, the photon number $|A|^{2}$ is dependent on the magnon number $|M|^2$.

\begin{figure}
\includegraphics[width=0.48\textwidth]{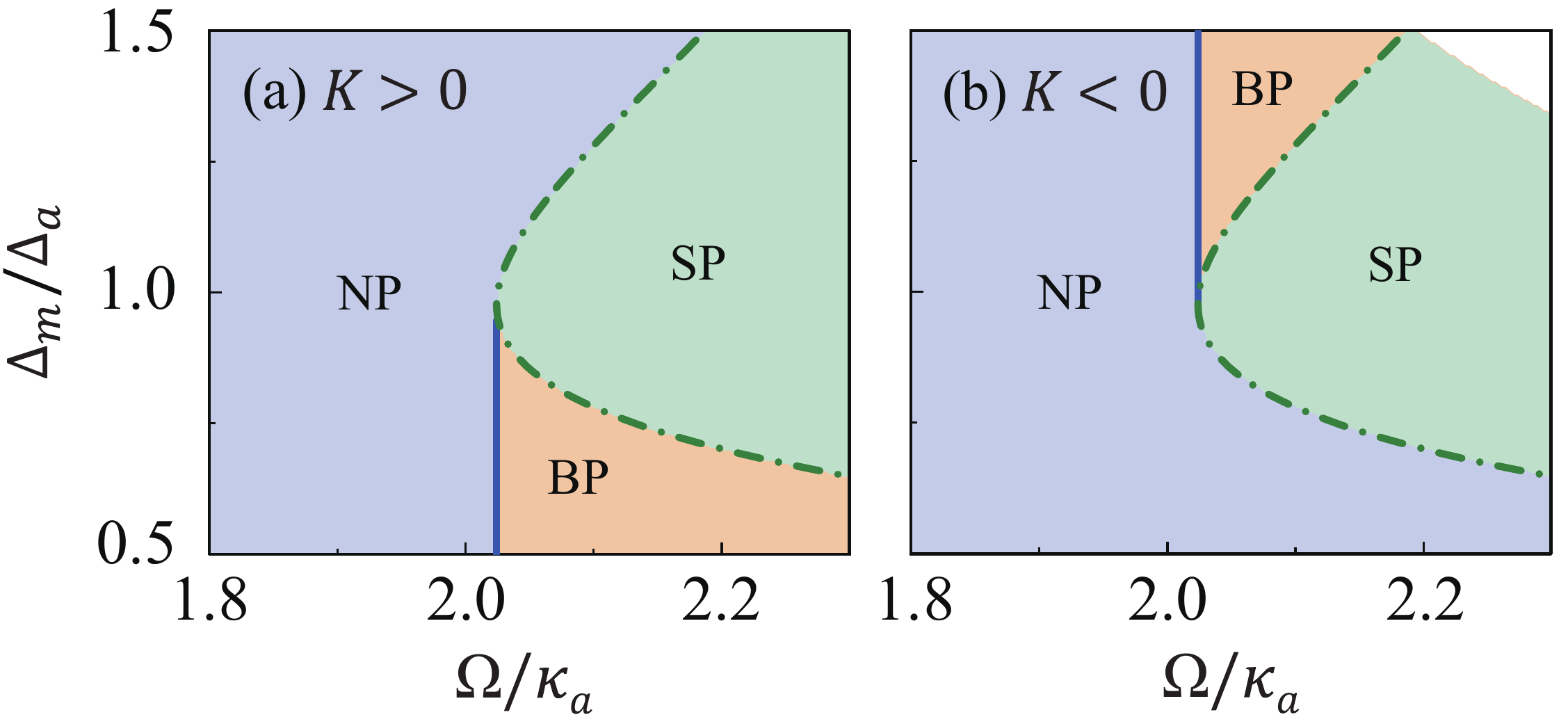}
\caption{Steady-state phase diagram of the cavity magnonic system versus the normalized drive strength $\Omega/\kappa_a$ and the detuning ratio $\Delta_m/\Delta_a$, where (a) $K > 0$ and (b) $K < 0$. NP, SP, and BP represent the normal phase, superradiant phase, and bistable phase, respectively. The white background in the upper-right corner of panel (b) indicates the unstable region. The vertical blue solid line marks the critical drive strength $\Omega=\Omega_1$, while the green dashed-dotted curve indicates the critical drive strength $\Omega=\Omega_2$. Other parameters are chosen to be $\Delta_{a}/\kappa_a=3$, $g_m/\kappa_a=2.4$, and $\gamma_m/\kappa_a=1$.}
\label{fig2}
\end{figure}

In this study, we choose the magnon number $|M|^2$ as the order parameter for the SQPT. The nonnegativity constraint on the magnon number $|M|^2$ restricts the valid solutions to $|M|^{2}_{\pm}\geq 0$. Table~\ref{Table1} summarizes the conditions for $|M|^{2}_+>0$ when $K>0$ and $|M|^{2}_->0$ when $K<0$, where we exclude the trivial solutions $|M|^{2}=|M|^{2}_-$ when $K>0$ and $|M|^{2}=|M|^{2}_+$ when $K<0$ due to their instability across all parameters (see Fig.~\ref{fig2} and related discussions). For $K>0$, the constraint $|M|^{2}_+ > 0$ requires $\Omega > \Omega_1$ when $\Delta_{m}/\Delta_{a}<\xi$ or $\Omega > \Omega_2$ for $\Delta_{m}/\Delta_{a}>\xi$, with the critical ratio $\xi$ defined as
\begin{equation}\label{xi}
\xi=\frac{2\gamma_m^{2}}{\sqrt{4(\Delta_a^{2}+\kappa_a^{2})\gamma_m^{2}
+(4\kappa_a\gamma_m+g_m^{2})g_m^{2}}-(2\kappa_a\gamma_m+g_m^{2})}.
\end{equation}
The critical drive strengths $\Omega_1$ and $\Omega_2$ are respectively given by
\begin{equation}\label{Omega1}
\Omega_1=\sqrt{\Delta_a^{2}+\kappa_a^2+(4\kappa_a\gamma_m+g_m^{2})g_m^{2}/4\gamma_m^{2}}-g_m^{2}/2\gamma_m,
\end{equation}
and
\begin{eqnarray}\label{Omega2}
\Omega_2=\sqrt{\Delta_a^{2}+\kappa_a^{2}-(2\Delta_a\Delta_m-2\kappa_a\gamma_m-g_m^{2})g_m^{2}/(\Delta_{m}^{2}+\gamma_m^{2})},~~~
\end{eqnarray}
which determine the boundaries between different phases in the steady-state phase diagram of the cavity magnonic system (see Fig.~\ref{fig2}). In particular, $\Omega_1=\Omega_2$ at $\Delta_{m}/\Delta_{a}=\xi$. For $K<0$, the constraint $|M|^{2}_- > 0$ is satisfied when $\Omega > \Omega_2$ for $\Delta_{m}/\Delta_{a}<\xi$ or $\Omega > \Omega_1$ for $\Delta_{m}/\Delta_{a}>\xi$.

\section{Nonreciprocal superradiant quantum phase transition}\label{NSQPT}

For examining the stability of the solutions for $|M|^2$ in Eq.~(\ref{solutions}), we need to investigate the fluctuations of the system. By introducing the quadrature operators $x_a=(\delta a^{\dag}+\delta a)/\sqrt{2}$, $y_a=i(\delta a^{\dag}-\delta a)/\sqrt{2}$, $x_m=(\delta m^{\dag}+\delta m)/\sqrt{2}$, and $y_m=i(\delta m^{\dag}-\delta m)/\sqrt{2}$, we recast Eq.~(\ref{fluctuations}) and its complex conjugate expression into the compact matrix form:
\begin{equation}\label{VQC}
\dot{u}(t)=\mathbf{\Lambda}\,u(t)+f_{\rm in}(t),
\end{equation}
where $u(t)=[x_a(t),\,y_a(t),\,x_m(t),\,y_m(t)]^T$ represents the vector of quadrature components, $f_{\rm in}(t)=[\sqrt{2\kappa_a}\,x^{\rm (in)}_a,\,\sqrt{2\kappa_a}\,y^{\rm (in)}_a,\,\sqrt{2\gamma_m}\,x^{\rm (in)}_m,\,\sqrt{2\gamma_m}\,y^{\rm (in)}_m]^T$ denotes the vector of noise quadratures, defined as $x^{\rm (in)}_a=(a_{\rm in}^{\dag}+a_{\rm in})/\sqrt{2}$, $y^{\rm (in)}_a=i(a_{\rm in}^{\dag}-a_{\rm in})/\sqrt{2}$, $x^{\rm (in)}_m=(m_{\rm in}^{\dag}+m_{\rm in})/\sqrt{2}$, and $y^{\rm (in)}_m=i(m_{\rm in}^{\dag}-m_{\rm in})/\sqrt{2}$, and the drift matrix $\Lambda$ is given by
\begin{equation}\label{matrix-Lambda}
\mathbf{\Lambda}=
\left(
  \begin{array}{cccc}
    -\kappa_a& \Delta_a-\Omega & 0 & g_m\\
    -\Delta_a-\Omega & -\kappa_a & -g_m & 0\\
    0 & g_m & -\gamma_m+{\rm Im}[F] & \widetilde{\Delta}_{m}-{\rm Re}[F]\\
    -g_m & 0 & -\widetilde{\Delta}_{m}-{\rm Re}[F] &-\gamma_m-{\rm Im}[F]\\
  \end{array}
\right).
\end{equation}
The stability of solutions $|M|_0^2$ and $|M|_\pm^2$ is governed by the spectral properties of the corresponding drift matrix $\mathbf{\Lambda}$~\cite{Gradshteyn80}. A given solution is stable only if all eigenvalues $\lambda_k$ of matrix $\mathbf{\Lambda}$ satisfy ${\rm Re}(\lambda_k) < 0$ for $k = 1, \,2, \,3, \,4$; otherwise, it is unstable.

\renewcommand\tabcolsep{11.9pt}
\begin{table}
	\centering
	\scriptsize
	\caption{Stable solutions (SSs) in each phase for both cases of $K > 0$ and $K < 0$.}
	\label{tab:notations}
    \small
	\begin{tabular}{ccc}
        \hline
		phase type  & \makecell[cc]{SSs for $K>0$} & \makecell[cc]{SSs for $K<0$}\\
		\hline
		\makecell[cc]{normal phase}  &\makecell[cc]{$|M|^2=0$}     & \makecell[cc]{$|M|^2=0$}\\
        \hline
		\makecell[cc]{superradiant phase}  & \makecell[cc]{$|M|^2=|M|^2_+$} &
                                 \makecell[cc]{$|M|^2=|M|^2_-$} \\
		\hline
        \makecell[cc]{bistable phase}  & \makecell[cc]{$|M|^2=0,\,|M|^2_+$} &
                                 \makecell[cc]{$|M|^2=0,\,|M|^2_-$} \\
		\hline
	\end{tabular}
\label{Table2}
\end{table}

Based on the stability analysis of the solutions in Eq.~(\ref{solutions}), we plot the steady-state phase diagram of the system as a function of the normalized drive strength $\Omega/\kappa_a$ and the detuning ratio $\Delta_m/\Delta_a$ in Fig.~\ref{fig2}. Note that the white background in the upper-right corner of Fig.~\ref{fig2}(b) indicates the unstable regime, where all three solutions for $|M|^2$ are unstable. Figures~\ref{fig2}(a) and \ref{fig2}(b) correspond to $K>0$ and $K<0$, respectively, which can be realized by aligning the bias magnetic field parallel to the [100] and [110] crystallographic axes of the YIG sphere~\cite{Wang18,Zhang19}. The difference between these two panels demonstrates the nonreciprocity of the SQPT. The phase diagram exhibits three distinct phases: normal phase (light blue area), superradiant phase (light green area) and bistable phase (light orange area), where the boundaries among them are determined by the vertical blue solid line [$\Omega = \Omega_1$ given in Eq.~(\ref{Omega1})] and the green dash-dotted curve [$\Omega = \Omega_2$ given in Eq.~(\ref{Omega2})]. Table~\ref{Table2} summarizes the stable solutions in each phase. In the normal phase, the solution $|M|^2 = 0$ is stable for both $K > 0$ and $K < 0$, corresponding to the vacuum state. In the superradiant phase, the solution $|M|^2 = |M|^2_+$ for $K > 0$ ($|M|^2 = |M|^2_-$ for $K < 0$) becomes stable, and the system exhibits macroscopic magnon excitations. In the bistable phase, both $|M|^2 = 0$ and $|M|^2 = |M|^2_+$ for $K > 0$ ($|M|^2 = 0$ and $|M|^2 = |M|^2_-$ for $K < 0$) are stable, and the system's initial state determines whether macroscopic magnon excitations occur~\cite{Zhang21-Chen}. For the parameters used in Fig.~\ref{fig2}, the critical ratio $\xi$ from Eq.~(\ref{xi}) is $\xi = 0.976$. When $\Delta_m/\Delta_a < \xi$, the boundary between normal phase and bistable phase (normal phase and superradiant phase) is given by $\Omega = \Omega_1$ ($\Omega = \Omega_2$) for $K > 0$ ($K < 0$). Conversely, when $\Delta_m/\Delta_a > \xi$, the boundary between normal phase and superradiant phase (normal phase and bistable phase) is given by $\Omega = \Omega_2$ ($\Omega = \Omega_1$) for $K > 0$ ($K < 0$). Regardless of the value of $\Delta_m/\Delta_a$ and the sign of $K$, the curve $\Omega = \Omega_2$ universally defines the boundary between bistable phase and superradiant phase.

\begin{figure}
\includegraphics[width=0.48\textwidth]{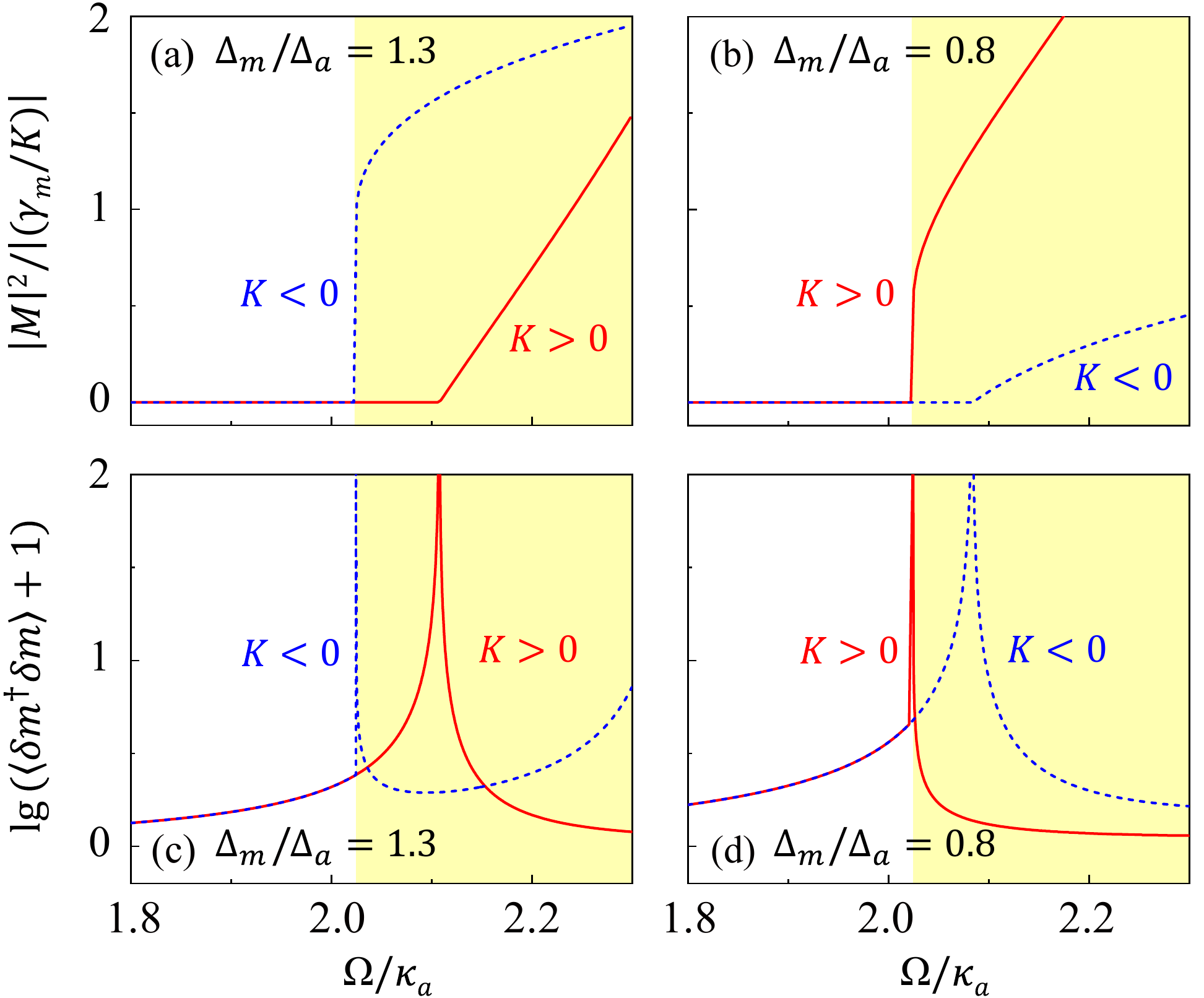}
\caption{(a),(b) Scaled steady-state magnon numbers $|M|^2/|(\gamma_m/K)|$ and (c),(d) magnon number fluctuations $\lg(\langle\delta m^\dagger \delta m \rangle + 1)$ versus the normalized  drive strength $\Omega/\kappa_a$, where $T=0$. Yellow shaded regions mark the nonreciprocal domains of SQPT, where $|M|^2(K>0) \neq |M|^2(K<0)$. The red solid curves correspond to $K > 0$, and the blue dashed curves correspond to $K < 0$. The detuning ratios are $\Delta_m/\Delta_a = 1.3$ and $\Delta_m/\Delta_a = 0.8$ in panels (a),(c) and (b),(d), respectively. Other parameters are the same as those in Fig.~\ref{fig2}.}
\label{fig3}
\end{figure}

In Figs.~\ref{fig3}(a), we display the scaled steady-state magnon number $|M|^2/|(\gamma_m/K)|$ as a function of the normalized drive strength $\Omega/\kappa_a$ for a fixed detuning ratio $\Delta_m/\Delta_a = 1.3$ ($> \xi = 0.976$). Yellow shaded regions indicate the nonreciprocal domains of SQPT, where $|M|^2(K>0) \neq |M|^2(K<0)$.
In the case of $K < 0$, the magnon number exhibits a discontinuous jump from zero to a finite value at the critical point $\Omega/\kappa_a = 2.025$ as the normalized drive strength $\Omega/\kappa_a$ increases across this threshold (blue dashed curve), signifying a first-order SQPT from the normal phase to the superradiant phase. However, for $K > 0$, the transition becomes second-order (i.e., a continuous SQPT), and the critical point shifts to $\Omega/\kappa_a = 2.108$ (red solid curve). The corresponding physical mechanisms are as follows. In the mean-field approximation, the Kerr term can be reduced to $\frac{K}{2}m^\dag m^\dag mm \approx \Delta_{\rm shift} m^\dag m$ with $\Delta_{\rm shift}=K\langle m^\dag m\rangle=K|M|^2$~\cite{Wang18}. This indicates that the MKE results in a blue shift $\Delta_{\rm shift}(K>0)$ [red shift $\Delta_{\rm shift}(K<0)$] in the magnon frequency when $K>0$ ($K<0$). For a weak parametric drive, the system remains stable even in the absence of magnon occupation [corresponding to $\Delta_{\rm shift}(K>0)=\Delta_{\rm shift}(K<0)=0$], and the sign of MKE coefficient bears no qualitative difference. However, under sufficiently strong parametric drive, the system tends to become unstable~\cite{Yuan21}. To counteract this tendency, the magnon occupation $\langle m^\dag m\rangle = |M|^2_+$ occurs, leading to a blue shift $\Delta_{\rm shift}(K>0)=K|M|^2_+$ for $K>0$ such that the system reaches a steady state at the shifted magnon frequency $\omega_m+\Delta_{\rm shift}(K>0)$. In contrast, when $K<0$, the system is generally unstable at the red-shifted magnon frequency $\omega_m+\Delta_{\rm shift}(K<0)$ with $\langle m^\dag m\rangle = |M|^2_+$. In fact, the steady state corresponds to the magnon occupation $\langle m^\dag m\rangle = |M|^2_-$ in the case of $K<0$. Figure~\ref{fig3}(b) shows $|M|^2/|(\gamma_m/K)|$ versus $\Omega/\kappa_a$ by taking $\Delta_m/\Delta_a = 0.8$ ($< \xi = 0.976$), where nonreciprocal SQPT is also evident. Unlike the case with $\Delta_m/\Delta_a > \xi$ [cf. Figs.~\ref{fig3}(b) and \ref{fig3}(a)], the system now undergoes a second-order SQPT from the normal phase to the superradiant phase at $\Omega/\kappa_a = 2.084$ for $K < 0$ (blue dashed curve), while the transition remains first-order at $\Omega/\kappa_a = 2.025$ if $K > 0$ (red solid curve).

To show more characteristics of the nonreciprocal SQPT, we examine the magnon number fluctuations $\langle\delta m^\dagger \delta m \rangle$ on top of the stable mean-field solutions for $|M|^2$. The quantum fluctuations of the cavity magnonic system are described by a $4\times4$ covariance matrix $\mathcal{V}(t)$, with elements defined as $\mathcal{V}_{\rm ij}(t)=\langle u_i(t) u_j(t')+u_j(t') u_i(t) \rangle / 2$ $(i,\,j = 1,\, 2,\, 3,\, 4)$. In our model, the magnon number fluctuations $\langle\delta m^\dagger \delta m \rangle= \left[ (V_{33} + V_{44}) - 1 \right] / 2$ are related to the diagonal elements of the steady-state covariance matrix $\mathcal{V}(+\infty)$~\cite{Zhu20}, which can be straightforwardly obtained by solving the so-called Lyapunov equation
\begin{equation}\label{}
\Lambda \mathcal{V} + \mathcal{V} \Lambda^T = -\mathcal{D},
\end{equation}
where the diffusion matrix $\mathcal{D}$, determined by $\mathcal{D}_{\rm ij} \delta(t - t') = \langle f_{{\rm in},i}(t)f_{{\rm in},j}(t') + f_{{\rm in},j}(t')f_{{\rm in},i}(t)\rangle / 2$, takes the form $\mathcal{D} = {\rm diag} \left[(2\bar{n}_a + 1)\kappa_a, (2\bar{n}_a + 1)\kappa_a, (2\bar{n}_m + 1)\gamma_m, (2\bar{n}_m + 1)\gamma_m  \right]$. For the nonreciprocal behavior of the SQPT shown in Figs.~\ref{fig3}(a) and \ref{fig3}(b), the corresponding magnon number fluctuations are presented in Figs.~\ref{fig3}(c) and \ref{fig3}(d), respectively. For clarity, we plot $\lg(\langle\delta m^\dagger \delta m \rangle + 1)$ as a function of the normalized drive strength $\Omega/\kappa_a$. Here, the red solid curves correspond to $K > 0$, and the blue dashed curves to $K < 0$. The magnon number fluctuations exhibit clear nonreciprocity, consistent with the nonreciprocal behavior of SQPT. In all cases, the fluctuations diverge near the critical point of the SQPT but approach zero far from the critical point [cf. Figs.~\ref{fig3}(a) and \ref{fig3}(b)].

\begin{figure}[tb]
\includegraphics[width=0.48\textwidth]{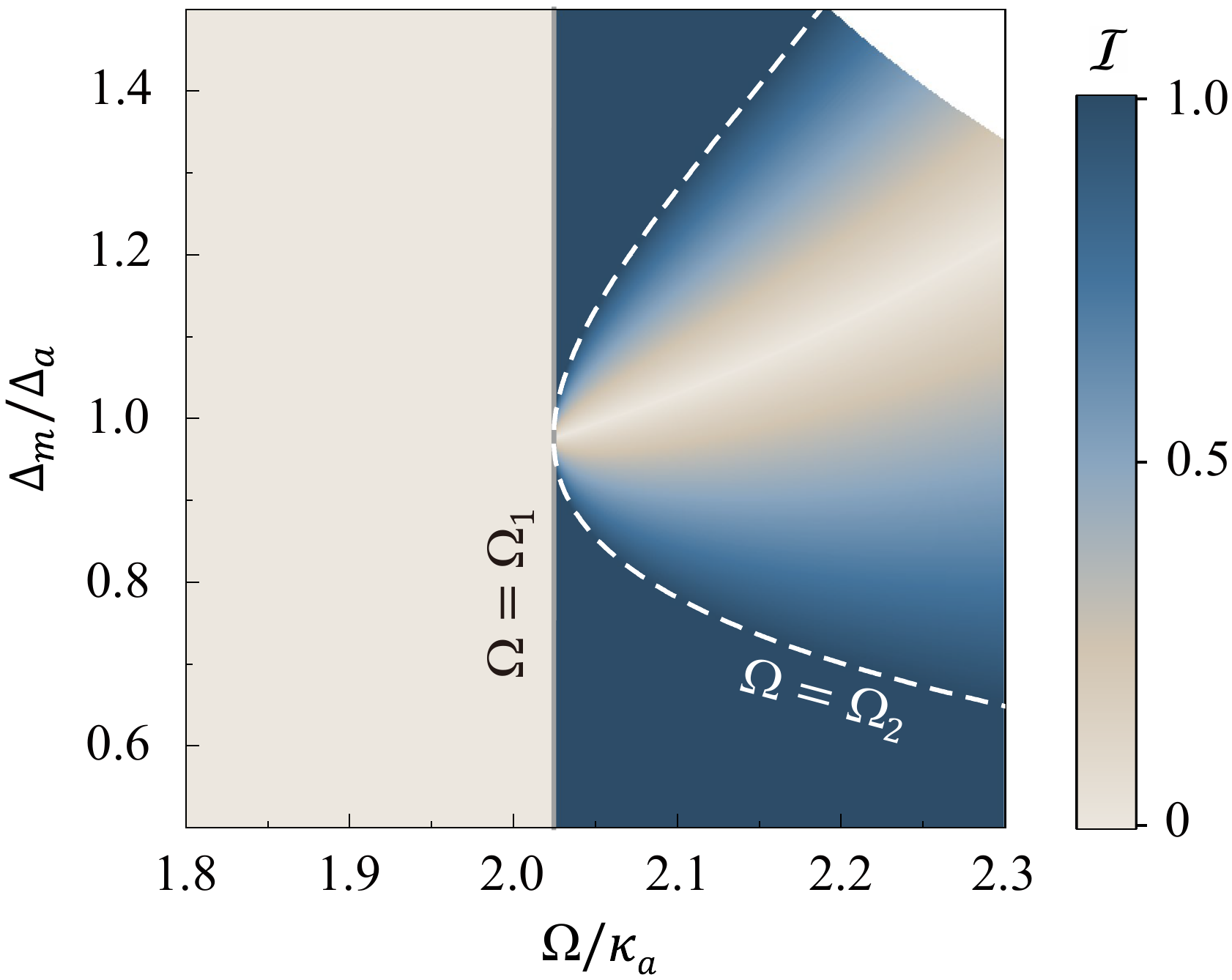}
\caption{Bidirectional contrast ratio $\mathcal{I}$ as a function of the normalized drive strength $\Omega/\kappa_a$ and the detuning ratio $\Delta_m/\Delta_a$. The white background in the upper-right corner corresponds to the unstable region in Fig.~\ref{fig2}(b). The vertical gray solid line marks the critical drive $\Omega = \Omega_1$, and the white dashed curve indicates $\Omega = \Omega_2$. Other parameters are the same as those in Fig.~\ref{fig2}.}
\label{fig4}
\end{figure}

Furthermore, we introduce the bidirectional contrast ratio:
\begin{equation}
\mathcal{I}= \left\{
\begin{aligned}
&0, &&  |M|^2(K>0) = |M|^2(K<0), \\
&\left|\frac{|M|^2(K>0)-|M|^2(K<0)}{|M|^2(K>0)+|M|^2(K<0)}\right|, &&  |M|^2(K>0) \neq |M|^2(K<0),
\end{aligned}
\right.
\end{equation}
which satisfies $0 \leq \mathcal{I} \leq 1$. This parameter quantitatively characterize the degree of nonreciprocity in the SQPT: a nonzero $\mathcal{I}$ signals the emergence of nonreciprocal behavior. The minimum value $\mathcal{I}_{\rm min}=0$ corresponds to a reciprocal SQPT, while the maximum value $\mathcal{I}_{\rm max}=1$ represents an ideal nonreciprocal SQPT. In Fig.~\ref{fig4}, we show the bidirectional contrast ratio $\mathcal{I}$ as a function of the normalized drive strength  $\Omega/\kappa_a$ and the detuning ratio $\Delta_m/\Delta_a$. The white region in the upper-right corner corresponds to the unstable regime identified in Fig.~\ref{fig2}(b). The vertical gray solid line (white dashed curve) marks the critical drive strength $\Omega = \Omega_1= 2.025\kappa_a$ ($\Omega = \Omega_2$). The bidirectional contrast ratio $\mathcal{I}$ attains its minimum value $\mathcal{I}_{\rm min}=0$ in the light beige region ($\Omega < \Omega_1$), where the system remains in the normal phase for both $K>0$ and $K<0$, indicating a reciprocal SQPT. Notably, two blue regions appear for $\Omega_1 < \Omega < \Omega_2$, where $\mathcal{I}$ reaches its maximum value $\mathcal{I}_{\rm max}=1$, corresponding to an ideal nonreciprocal SQPT. In the lower (upper) blue region, the system is in the superradiant phase (normal phase) for $K>0$ but remains in the normal phase (superradiant phase) for $K<0$. In the remaining region ($\Omega > \Omega_2$), the bidirectional contrast ratio lies between 0 and 1 (i.e., $0<\mathcal{I}<1$). Here, the system is in the superradiant phase for both $K>0$ and $K<0$, but the order parameters differ: $|M|^2(K>0)=|M|^2_+ \neq |M|^2(K<0)=|M|^2_-$.

\begin{figure}
\includegraphics[width=0.48\textwidth]{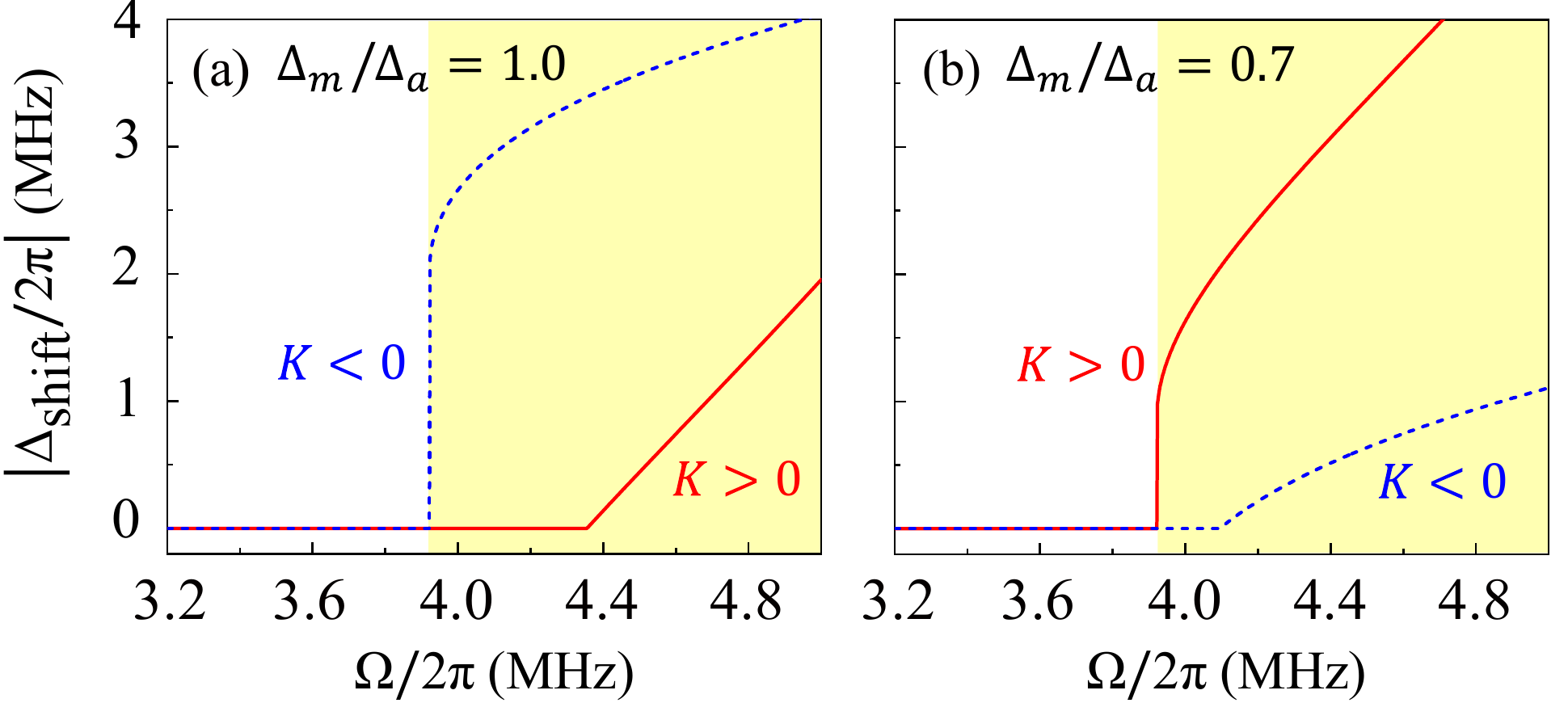}
\caption{MKE-induced magnon frequency shift $\Delta_{\rm shift}$ versus the drive strength $\Omega$ for (a) $\Delta_m/2\pi=9.77$~MHz and (b) $\Delta_m/2\pi=6.84$~MHz. The red solid curves correspond to $K/2\pi=4.1$~nHz, and the blue dashed curves correspond to $K/2\pi=-3.3$~nHz. Other parameters are chosen to be $\kappa_a/2\pi=2.04$~MHz, $\gamma_m/2\pi=1.49$~MHz, $g_m/2\pi=8.17$~MHz, $T=10$~mK, $\omega_a/2\pi=4.572$~GHz, and $\Delta_a/2\pi=9.77$~MHz.}
\label{fig5}
\end{figure}

\section{Experimental feasibility}\label{experimental}

In this section, we will assess the feasibility of the physical implementation of the proposed scheme. For implementing the proposed scheme, we consider a cavity magnonic system comprising a parametrically driven coplanar waveguide resonator (CWR) coupled to a YIG sphere. The parametric drive on the CWR is supplied by a flux-driven Josephson parametric amplifier (JPA). Experimentally, strong coupling between the CWR mode and the magnon mode in a 250-$\mu$m-diameter YIG sphere has been demonstrated in Refs.~\cite{Morris17,Li22,Song25}, where the corresponding MKE coefficient is $K/2\pi=4.1$~nHz ($-3.3$~nHz)~ for the bias magnetic field aligned along the crystallographic axis [100] ([110])~\cite{Zhang19}. In these setups, the frequencies of CWR and magnon modes are typically around 5~GHz, while their decay rates are on the order of 1~MHz~\cite{Morris17,Li22,Song25}. Moreover, the magnon frequency can be readily tuned by adjusting the bias magnetic field. As demonstrated in Ref.~\cite{Wang18}, the bistability of cavity magnon polaritons induced by MKE has been observed in the two cases of aligning the bias magnetic field along the [100] and [110] crystallographic axes of the YIG sphere, corresponding to positive and negative MKE coefficients, respectively. The action of the flux-driven JPA on the CWR is well described by a parametric drive Hamiltonian [cf.~the last term in Eq.~(\ref{A1})]~\cite{Yamamoto08,Zhong13}, where the drive strength (e.g., tunable from 0 to 6~MHz~\cite{Krantz16}) and frequency can be precisely controlled via the external flux drive applied to the JPA.

\begin{figure}
\includegraphics[width=0.48\textwidth]{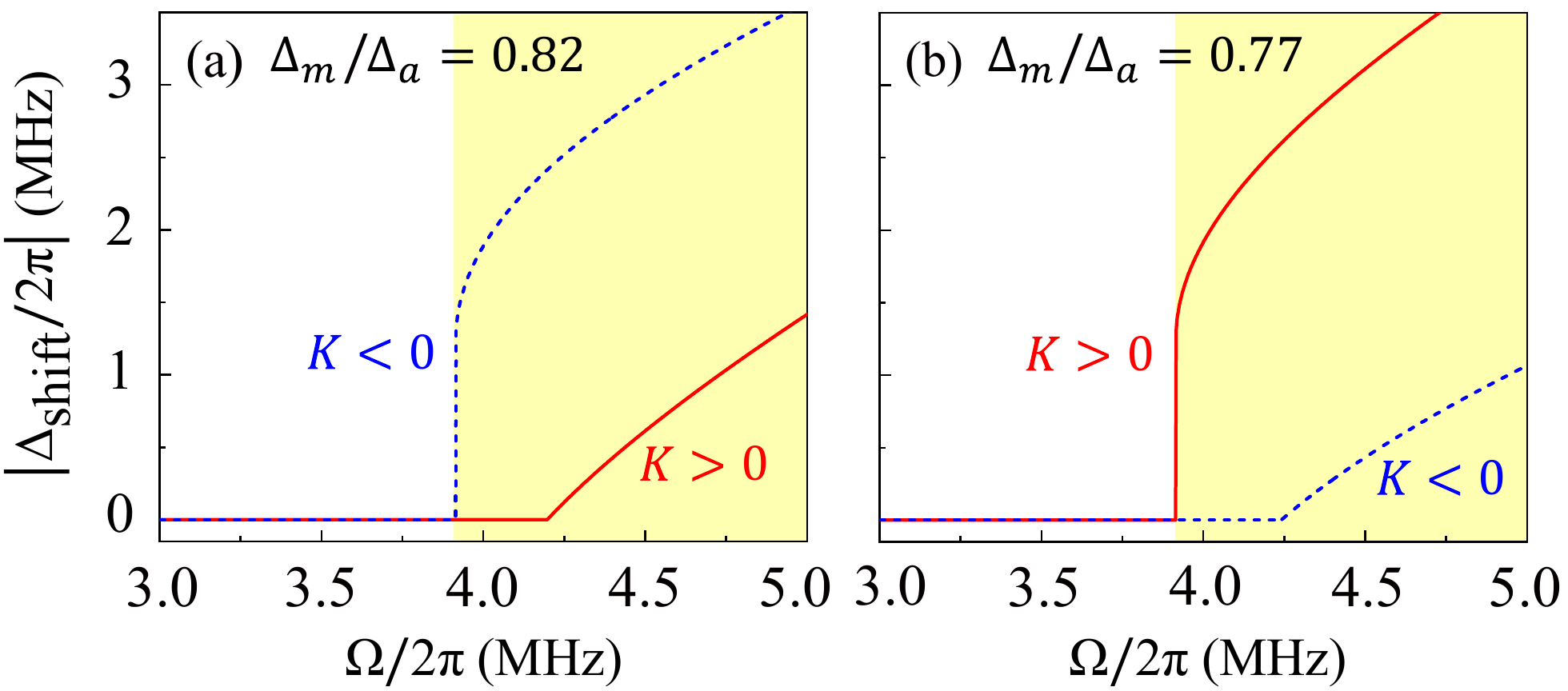}
\caption{MKE-induced magnon frequency shift $\Delta_{\rm shift}$ versus the drive strength $\Omega$ for (a) $\Delta_m/2\pi=37$~MHz and (b) $\Delta_m/2\pi=34.5$~MHz, with $g_m/2\pi=40$~MHz and $\Delta_a/2\pi=45$~MHz. The red solid curves correspond to $K/2\pi=4.1$~nHz, and the blue dashed curves correspond to $K/2\pi=-3.3$~nHz. Other parameters are the same as those in Fig.~\ref{fig5}.}
\label{fig6}
\end{figure}

In the experiment, the nonreciprocal SQPT can be detected by measuring the MKE-induced magnon frequency shift $\Delta_{\rm shift}=K|M|^2$ via the transmission spectroscopy of the cavity~\cite{Wang16,Wang18}. With experimentally feasible parameters $\kappa_a/2\pi=2.04$~MHz, $\gamma_m/2\pi=1.49$~MHz, $g_m/2\pi=8.17$~MHz, $K/2\pi=4.1$~nHz ($-3.3$~nHz), $T=10$~mK, $\omega_a/2\pi=4.572$~GHz~\cite{Morris17}, and $\Omega/2\pi < 6$~MHz~\cite{Krantz16}, we numerically simulate the magnon frequency shift $\Delta_{\rm shift}$ versus the drive strength in Fig.~\ref{fig5}, which clearly shows the nonreciprocal SQPT. In fact, our theoretical proposal does not place stringent constraints on the cavity-magnon coupling strength $g_m$, provided that the system remains within the strong-coupling regime (i.e., $ \{\kappa_a,\, \gamma_m\}_{\rm max} \ll g_m \ll \{\omega_a,\, \omega_m\}_{\rm min}$). For a given set of parameters $\{\kappa_a, \gamma_m,g_m\}$, the nonreciprocal SQPT can be observed in the parameter region where $\Delta_m \approx \xi \Delta_a$ and $\Omega \approx \kappa_a+\gamma_m$ by choosing $\Delta_a \approx g_m$ (see Appendix~\ref{Appendix-B}). For example, if increasing the cavity-magnon coupling to $g_m/2\pi=40$~MHz while keeping other system parameters unchanged, the nonreciprocal SQPT can appears in the region  $\Delta_m/2\pi \approx 35.8$~MHz and $\Omega/2\pi \approx 3.53$~MHz by choosing $\Delta_a/2\pi=45$~MHz (see Fig.~\ref{fig6}).

It should be noted that we employ a spherical YIG sample. This choice is motivated by the fact that spherical YIG samples host magnon modes with low dissipation rates ($1 \sim 2$~MHz)~\cite{Morris17,Li22,Song25}. This helps keep the required parametric-drive strength for inducing the nonreciprocal SQPT within a practical range, since the critical drive is given by $\Omega \approx \kappa_a+\gamma_m$ (see Appendix~\ref{Appendix-B}). Experimentally, the strong coupling between a magnon mode in a thin-film magnetic material and a CRW has also been demonstrated, where the decay rate of the magnon mode is on the order of 50 \textemdash 180~MHz~\cite{Huebl13,Hou19,Li19}. In such thin-film magnetic materials, both magnetocrystalline anisotropy and shape anisotropy can also lead to a direction-dependent MKE coefficient. If future fabrication advances enable further reduction of magnon dissipation in thin-film magnetic materials, our scheme could be implemented in on-chip ferromagnet-superconductor thin-film devices~\cite{Huebl13,Hou19,Li19}. However, our scheme is no longer applicable when the cavity-magnon system enters the ultrastrong-coupling regime~\cite{Ghirri23}. This limitation arises because the scheme neglects the counter-rotating wave terms [cf.~Eq.~(\ref{NHH})], which restricts its validity to the strong-coupling regime.

\section{Discussion and conclusions}\label{conclusions}

In our analysis, we employ the mean-field approximation [cf.~Eqs.~(\ref{Langevin})\textemdash(\ref{fluctuations})], which is justified in the case of $|K| \ll \gamma_m$~\cite{HHZhang21}. If the condition $|K| \ll \gamma_m$ cannot be satisfied, the mean-field approximation fails, and the SQPT may be obscured. For a 0.25-mm-diameter YIG sphere, the MKE coefficient $K$ is on the order of several nHz~\cite{Zhang19}, which ensures that the mean-field approximation in our analysis is valid. Although the MKE is very weak, the MKE-induced magnon frequency shift $\Delta_{\rm shift}=K|M|^2$ is on the order of several MHz (see Figs.~\ref{fig5} and \ref{fig6}), corresponding to a magnon population on the order of $|\gamma_m/K| \sim 10^{16}$. This is consistent with experimental observations of magnon-polariton bistability in Ref.~\cite{Wang18}, where MKE-induced $\Delta_{\rm shift}$ was observed to reach up to 80~MHz by employing a coherent field to pump a 1-mm-diameter YIG sphere. Under a coherent-field drive [corresponding to replacing $\frac{\Omega}{2}(a^{\dag}a^{\dag}+aa)$ in Eq.~(\ref{NHH}) with $\frac{\Omega}{2}(a^{\dag}+a)$], the system Hamiltonian takes the form~\cite{Wang18}
\begin{eqnarray}\label{BH}
H_{\rm bista}&=&(\Delta_{a}-i\kappa_a)a^{\dag}a+(\Delta_{m}-i\gamma_m)m^{\dag}m+\frac{K}{2}m^{\dag}m^{\dag}mm\nonumber\\
           & & +g_m(a^{\dag}m+am^{\dag})+\frac{\Omega}{2}(a^{\dag}+a),
\end{eqnarray}
which no longer preserves parity symmetry, i.e., $[H_{\rm bista},\Pi] \neq 0$. Although both the SQPT and the bistability induced by NKE can be experimentally probed via the magnon frequency shift $\Delta_{\rm shift} = K|M|^{2}$, they are fundamentally different. In the SQPT, $\Delta_{\rm shift}$ stays zero below a critical drive threshold, while exhibiting a sharp onset above it [cf.~Figs.~\ref{fig5} and~\ref{fig6}]. Conversely, for bistability, $\Delta_{\rm shift}$ remains finite at any drive strength~\cite{Wang18}.

In principle, a stronger MKE can be introduced by coupling an ancillary qubit to the magnon mode~\cite{Lachance-Quirion17}. When the ancillary qubit is strongly coupled to the magnon mode and its detuning $\Delta_{\rm mq}$ from the magnon mode is much larger than the magnon-qubit coupling strength $g_{\rm mq}$ (i.e., $\{\gamma_m,\,\gamma_q\}_{\rm max} \ll g_{\rm mq} \ll \Delta_{\rm mq}$ with qubit decay rate $\gamma_q$), we can derive the reduced magnon Hamiltonian as $H_m=\omega_m m^\dag m+\frac{K_m}{2} m^\dag m^\dag mm$ with $K_m=2g_{\rm mq}^4/\Delta_{\rm mq}^3$ by eliminating the qubit degree of freedom, where system dissipation is negligible in the derivation. However, this qubit-induced MKE cannot trigger the specific SQPT studied in our work. This limitation arises from two fundamental constraints. First, the ancillary qubit is indirectly coupled to the magnon mode through an auxiliary cavity field, which substantially increases the complexity of the experimental implementation~\cite{Lachance-Quirion17}. More critically, the reduced Hamiltonian $H_m$ is valid only under the low-population condition $\langle m^\dag m\rangle \ll \langle m^\dag m\rangle_{\rm crit}=\Delta_{\rm mq}^2/4g_{\rm mq}^2$~\cite{Boissonneault09}, which cannot be met in our proposal. When the SQPT occurs, the magnon population is on the order of $\gamma_m/K_m=\Delta_{\rm mq}^3\gamma_m/(2g_{\rm mq}^4)$, which is comparable to the critical population $\langle m^\dag m\rangle_{\rm crit}$. Using typical parameters from Ref.~\cite{Lachance-Quirion17}, for example, $\langle m^\dag m\rangle_{\rm crit} = 4.6$, whereas $\langle m^\dag m\rangle \sim \gamma_m/K_m = 6.5$.

In conclusion, we have demonstrated that the MKE in a cavity magnonic system offers a viable pathway to achieve a nonreciprocal SQPT~\cite{Fruchart21,Chiacchio23,Zhu24,Xu24}. The proposed setup, comprising a YIG sphere coupled to a microwave cavity, exhibits markedly different phase diagrams under positive and negative MKE coefficients---controlled by orienting the bias magnetic field along different crystallographic axes of the YIG sphere~\cite{Wang18,Zhang19}. These distinct phase boundaries directly lead to nonreciprocal critical behavior, as confirmed by analyzing the magnon occupation and its fluctuations around the transition. The introduction of a bidirectional contrast ratio further allows quantitative characterization of the nonreciprocity.

Our results establish cavity magnonics as a promising platform for exploring nonreciprocal SQPT. The nonreciprocal SQPT is not only of fundamental interest in studying nonreciprocal collective phenomena~\cite{Fruchart21}, but also brings potential applications in, e.g.,nonreciprocal signal transmission~\cite{Kong19,Wang19-Rao} and unidirectional quantum sensing~\cite{Garbe20,Chu21}. In the nonreciprocal domains of SQPT, the magnon mode exhibits distinct frequency shifts for positive and negative MKE coefficients, i.e., $\Delta_{\rm shift}(K>0) \neq \Delta_{\rm shift}(K<0)$ (cf.~Figs.~\ref{fig5} and \ref{fig6}). This spectral distinction implies different underlying energy spectra for $K>0$ and $K<0$, suggesting the feasibility of controlling signal transmission through nonreciprocal SQPT~\cite{Kong19,Wang19-Rao}. Moreover, integrating nonreciprocal SQPT with SQPT-based quantum sensing could enable unidirectional quantum sensing, which would allow direct detection of signal directionality~\cite{Garbe20,Chu21}. In contrast to conventional reciprocal quantum sensing, such a unidirectional sensing scheme naturally suppresses noise originating from directions outside the detection channel.

\section*{Acknowledgments}

This work is supported by the National Natural Science Foundation of China (Grants No.~12205069, No.~12204139, and No.~12504343), the HZNU scientific research and innovation team project (Grant No. TD2025003), the Hangzhou Leading Youth Innovation and Entrepreneurship Team project (Grant No. TD2024005), and the Zhejiang Provincial Natural Science Foundation of China (Grant No.~LQN25A040019). Y.H.K. is supported by the National Key Research and Development Program of China (Grant No. 2024YFA1408900). W.X. is supported by the Natural Science Foundation of Zhejiang Province (Grant No. LY24A040004), the Zhejiang Province Key R\&D Program of China (Grant No. 2025C01028), and the Shenzhen International Quantum Academy (Grant No. SIQA2024KFKT010).

\appendix

\section{Derivation of the effective non-Hermitian Hamiltonian in Eq.~(\ref{NHH})}\label{Appendix-A}

As shown in Fig.~\ref{fig1}, the magnon mode in the YIG sphere is coupled to the cavity mode. Including both the parametric drive and the MKE, the total Hamiltonian of the system is given by
\begin{eqnarray}\label{A1}
H &=& \omega_a a^{\dag}a+\omega_m m^{\dag}m+\frac{K}{2}m^{\dag}m^{\dag}mm+g_m (a^{\dag}m+am^{\dag})\nonumber\\
  & & +\frac{\Omega}{2}(a^{\dag}a^{\dag}e^{-i\omega_{d}t}+aae^{i\omega_{d}t}),
\end{eqnarray}
where $a^{\dag}$ and $m^{\dag}$ ($a$ and $m$) are the creation (annihilation) operators for the cavity and magnon modes with frequencies $\omega_a$ and $\omega_m$, respectively, $K$ is the MKE coefficient, $g_m$ is the coupling strength between cavity and magnon modes, and $\Omega$ ($\omega_{d}$) is the strength (frequency) of parametric drive. In the rotating frame at $\omega_{d}/2$, the Hamiltonian in Eq.~(\ref{A1}) becomes
\begin{eqnarray}\label{A2}
H&=&\Delta_{a}a^{\dag}a+\Delta_{m}m^{\dag}m+\frac{K}{2}m^{\dag}m^{\dag}mm+g_m (a^{\dag}m+am^{\dag})\nonumber\\
 & &+\frac{\Omega}{2}(a^{\dag}a^{\dag}+aa),
\end{eqnarray}
where $\Delta_a=\omega_a-\omega_d/2$ and $\Delta_m=\omega_m-\omega_d/2$ represent the detunings of the cavity and magnon modes from the parametric drive, respectively. With the Heisenberg-Langevin approach~\cite{Walls94}, the quantum dynamics of the system is governed by
\begin{eqnarray}\label{A3}
\dot{a}&=&-i(\Delta_{a}-i\kappa_a)a-ig_m m-i\Omega a^{\dag}+\sqrt{2\kappa_a}\,a_{\rm in},\nonumber\\
\dot{m}&=&-i(\Delta_{m}-i\gamma_m)m-iKm^{\dag}mm-ig_m a+\sqrt{2\gamma_m}\,m_{\rm in},~~~
\end{eqnarray}
which is Eq.~(\ref{Langevin}) in the main text. Here, $\kappa_a$ ($\gamma_m$) denotes the decay rate of the cavity mode (magnon mode), and $a_{\rm in}$ ($m_{\rm in}$) is the corresponding input noise operator. Rewriting Eq.~(\ref{A3}) in the form
\begin{eqnarray}\label{A4}
\dot{a}&=&-i[a,H_{\rm eff}]+\sqrt{2\kappa_a}\,a_{\rm in},\nonumber\\
\dot{m}&=&-i[m,H_{\rm eff}]+\sqrt{2\gamma_m}\,m_{\rm in},
\end{eqnarray}
yields the effective non-Hermitian Hamiltonian $H_{\rm eff}$ presented in Eq.~(\ref{NHH}).

\section{A guide to choosing system parameters for experimental implementation}\label{Appendix-B}

The successful experimental implementation of this theoretical proposal hinges on the appropriate selection of system parameters. The critical drive strengths given in Eqs.~(\ref{Omega1}) and (\ref{Omega2}) can be recast as
\begin{equation}\label{Omega1B}
\Omega_1=\frac{1}{2\gamma_m}\left[\sqrt{(g_m^2+2\kappa_a\gamma_m)^2+4\Delta_a^2 \gamma_m^2}-g_m^{2}\right],
\end{equation}
and
\begin{eqnarray}\label{Omega2B}
\Omega_2=\sqrt{\frac{(\Delta_a\Delta_m-g_m^2 - \kappa_a\gamma_m)^2+(\Delta_a\gamma_m+\Delta_m\kappa_a)^2}{\Delta_m^2+\gamma_m^2}}
\,.
\end{eqnarray}
A direct inspection shows that both $\Omega_1$ and $\Omega_2$ remain positive for any value of the coupling strength $g_m$. Consequently, the proposed scheme does not impose a stringent requirement on $g_m$. Since cavity-magnonic systems are typically operated experimentally in the strong-coupling regime (i.e., $ \{\kappa_a,\, \gamma_m\}_{\rm max} \ll g_m \ll \{\omega_a,\, \omega_m\}_{\rm min}$)~\cite{Morris17,Li22,Song25}, the following discussion will focus on the selection of experimental parameters for such strongly coupled systems.

For a given set of parameters $\{\kappa_a, \gamma_m,g_m\}$, the cavity-mode detuning is set to $\Delta_a \approx g_m$. This condition can be readily satisfied by tuning the frequency of the parametric drive. Once $\Delta_a$ is fixed, the critical ratio
$\xi$ ($\approx 1$) defined in Eq.~(\ref{xi}) is determined, which in turn sets the magnon detuning via the relation $\Delta_m \approx \xi \Delta_a$. The magnon detuning $\Delta_m$ can be experimentally controlled by adjusting the bias magnetic field applied to the YIG sphere. With $\Delta_m/\Delta_a=\xi$, the critical drive strengths are obtained as $\Omega_1=\Omega_2 \approx \kappa_a+\gamma_m$. In summary, by choosing $\Delta_a \approx g_m$ for a fixed set of $\{\kappa_a, \gamma_m,g_m\}$, the nonreciprocal SQPT can be observed in the parameter region where $\Delta_m \approx \xi \Delta_a$ and $\Omega \approx \kappa_a+\gamma_m$. This result clearly shows that the critical drive strengths $\Omega_1$ and $\Omega_2$ required to induce the SQPT exhibit only a weak dependence on the cavity-magnon coupling strength $g_m$. To illustrate with a concrete example, we take the parameters $\kappa_a/2\pi=2.04$~MHz, $\gamma_m/2\pi=1.49$~MHz, and $g_m/2\pi=8.17$~MHz from Ref.~\cite{Morris17}. Setting $\Delta_a/2\pi=9.77$~MHz, the nonreciprocal SQPT is predicted to occur in the region where $\Delta_m/\Delta_a \approx 0.792$ and $\Omega/2\pi \approx 3.53$~MHz (see Fig.~\ref{fig5}). If the coupling strength is increased to $g_m/2\pi=40$~MHz while keeping $\kappa_a$ and $\gamma_m$ unchanged, and with $\Delta_a/2\pi=45$~MHz, the nonreciprocal SQPT appears in the region where $\Delta_m/\Delta_a \approx 0.795$ and $\Omega/2\pi \approx 3.53$~MHz (see Fig.~\ref{fig6}).

\end{document}